\documentstyle[12pt,aaspp,psfig]{article}
\singlespace
\tighten
\onecolumn
\def \be{\begin{equation}}
\def \ee{\end{equation}}
\def \acknowledgments{\vskip 3ex plus .8ex minus .4ex}

\def\lesssim{\mathrel{\hbox{\rlap{\hbox{\lower4pt\hbox{$\sim$}}}\hbox{$<$}}}}
\def\gtrsim{\mathrel{\hbox{\rlap{\hbox{\lower4pt\hbox{$\sim$}}}\hbox{$>$}}}}
\def \units{\hspace{1cm}}
\def \and {\&}

\def \cm{~{\rm cm}}
\def \solarmass{~\! M_\odot}

\def \pc{~{\rm pc}}

\def \km{~{\rm km}}

\def \Gyr{~{\rm Gyr}}

\def \persec{~{\rm s}^{-1}}

\def \period{\hspace{1cm} . }

\def \eg{e.g.~}
\def \etal{et.~$\!$al.~$\!$}
\def\mstar{ m_{*} }
\def\rstar{ r_{*} }

\def\spa{~\!}

\begin{document}

\title{Dynamical Constraints On Alternatives To Supermassive Black \\
       Holes In Galactic Nuclei}

\vspace{0.8in}
\author{Eyal Maoz}
\vspace{0.4in}
\affil{Astronomy Department, University of California, Berkeley, CA 94720}

\vspace{0.7in}
\centerline{$\dagger$ \it To appear in the Astrophysical Journal (Letters)}

\newpage
\begin{abstract}
The compelling dynamical evidence for massive dark objects in galactic 
nuclei does not uniquely imply massive black holes (BHs). 
To argue convincingly that these objects are BHs we must
rule out alternatives to a BH, and the alternative to a point mass is
a cluster of some sort of nonluminous objects, such as a cluster of brown
dwarfs or stellar remnants.

We use simple physical considerations 
to derive the maximum possible lifetime of a dark cluster 
which may consist of any plausible form of non-luminous gravitating objects -- 
from brown dwarfs and very low-mass objects of cosmic composition, to 
white dwarfs, neutron stars, and black holes. 
The lower this limit relative to the galaxy age, the more
implausible is the cluster hypothesis, thus arguing for a point mass.
A cluster with a lifetime much shorter than $10\Gyr$ is unacceptable,
since observing it at the present epoch would be highly improbable.

Since the goal is to rule out a dark cluster by showing that its lifetime
must be very short, we make the most generous assumptions possible under the 
observational constraints to allow for its survival.
We find that the lifetime of such an hypothetical cluster must be much 
shorter than the galaxy age only in the cases of NGC 4258 and our Galaxy, 
thus strongly arguing for a point mass.  In all other galaxies, the case
of a massive BH, although compelling, is not yet watertight.
We also note that there are two exotic alternatives
to a massive BH that cannot be ruled out even in the cases of NGC~4258 and 
the Galaxy: clusters of elementary particles (e.g.~bosons),
and clusters of very low-mass ($\lesssim\!0.04\solarmass$) BHs.
We point out, however, serious difficulties with these 
alternatives, and argue that they are highly implausible.

\end{abstract}


\newpage
\section{INTRODUCTION}

There is now compelling evidence for massive dark objects at the
centers of severals galaxies (\eg Kormendy \& Richstone 1995, hereafter KR95). 
The possibility  that these are 
black holes (BHs) fits well into the picture where quasars and AGNs are
powered by accretion onto a massive BH, so that dead quasar engines should be
hiding in many nearby galaxies.
However, as emphasized by KR95, 
the massive BH picture has become a paradigm, which is a dangerous situation
since it is easy to believe that we have proved what we expect to find.
High $M/L$ ratios and gas velocities of 
order $\sim\!10^3 \km\persec$ in galactic centers do not uniquely
imply massive BHs.  
To argue convincingly that these objects are BHs we must
rule out alternatives to a BH, and the alternative to a point mass is
a cluster of some sort of nonluminous stars, such as a cluster of
stellar remnants, brown dwarfs or very low-mass objects.  
We should not exclude these possibilities even if the formation of such 
clusters might seem implausible. After all, the physical conditions in galactic 
centers are much different from those in the solar vicinity, and our 
understanding of star formation is still limited.
We note that the existence of $10^{6}\hbox{--}10^{9.5} \solarmass$ BHs is 
still the more exotic of the above interpretations, and its
acceptance requires extraordinary evidence. 

Dynamical data alone cannot provide a rigorous proof of a massive BH, unless
relativistic velocities are detected at a few Schwarzschild radii.
They can, however, provide an upper limit to the lifetime of an hypothetical
cluster. The lower this limit relative to the galaxy age, the more
implausible is the cluster hypothesis, thus arguing for a point mass.
A cluster with a lifetime much shorter than $\sim\!10$ Gyr is unacceptable 
because observing it at the present epoch would be highly improbable.
In \S2 we use simple physical considerations to derive the maximum
possible lifetime of a dark cluster which may consist of any plausible form 
of non-luminous gravitating objects -- from brown dwarfs and very low-mass
objects of cosmic composition, to stellar remnants.  We describe a
criterion for ruling out such alternatives to a massive BH, and apply it to 
observed cases. 
In \S3 we discuss two exotic possibilities that cannot be excluded -- clusters 
of elementary particles and very low-mass BHs, 
and argue that they are highly implausible.

\section{LIMITS TO THE LIFETIME OF A DARK CLUSTER}

\subsection{Structure and Composition of Dark Clusters}

Since the goal is to rule out a dark cluster by showing that its lifetime
must be very short, the cluster must be chosen in the most generous way 
to allow for its survival.  
Thus, we shall assume that (i) the cluster is of lowest
possible concentration, which would minimize the stellar collision rate; 
(ii) the cluster consists of equal-mass objects, since otherwise mass
segregation would accelerate its evolution; and 
(iii) the objects comprising the cluster have zero temperature, thus
having the smallest possible radii at a given mass, which would minimize
the stellar collision rate. 
We also assume for simplicity an isotropic velocity distribution. 

The least centrally concentrated model for a cluster with a given mass and 
half-mass radius is a nearly uniform core with the steepest possible
density falloff at larger distances.
Thus, we shall assume a Plummer model for the cluster structure, which has 
the steepest asymptotic density profile observed in any astrophysical system,
\be
\rho(r) = \rho_0 \left({1 + {r^2 \over r_c^2} }\right)^{-5/2} , 
\label{def_Plummer} \ee
where $\rho_0\!=\!3M/4\pi r_c^3$ is the central density, $r_c$ is the core
radius, and $M$ is the cluster's total mass.  It is useful to 
replace the two cluster parameters, $\rho_0$ and $r_c$, by the cluster's 
half mass, $M_h\!\equiv\!M/2$, and its 
half-mass density, $\rho_h$, which is the
mean density within the cluster's half-mass radius, $R_h$. 
In the case of a Plummer model we have $R_h\!=\!1.3 r_c$ and
$\rho_0\!=\!4.4\rho_h$.

We shall examine all plausible classes of non-luminous objects which
may comprise a dark cluster: (i) BHs with $\mstar\!\gtrsim\!3\solarmass$;
(ii) neutron stars with $1.4\!\lesssim\!\mstar\!\lesssim\!3\solarmass$;
(iii) very low-mass objects with
$\mstar\!\lesssim\!3\!\times\!10^{-3} \solarmass$, where the gravitational
forces are small compared with the electrostatic forces, as in planets;  and
(iv) objects with masses in the range $3\!\times\!10^{-3} \!\lesssim \mstar \!
\lesssim 1.4\solarmass$, where gravity is balanced by electron degeneracy
pressure. These include brown dwarfs, which are hydrogen rich and thus 
have masses up to the H-burning mass limit ($\simeq\!0.09\solarmass$),
and white dwarfs which can have any mass within this range.
A cluster lifetime depends on the mass and size of the objects comprising it. 
For white dwarfs at zero temperature we shall assume the mass-radius relation 
derived by Nauenberg (1972),
\be  \rstar(\mstar)   ~=~ { 1.57\!\times\!10^9 \over \mu}~
{ [1 - (\mstar/M_3)^{4/3}]^{1/2}  \over (\mstar/M_3)^{1/2}  } ~\cm,  
\hspace{0.4truein} (3\!\times\!10^{-3}\!\lesssim\mstar\!\lesssim\!1.4)
\ee
where $\mu$ is the mean molecular weight, and
$M_3\!\equiv\! 5.816 \mu^{-2}~\!M_\odot$ is the Chandrasekhar's limit. 
For cold brown dwarfs and lower mass objects of cosmic composition, 
we shall assume the mass-radius relation derived by 
Zapolsky and Salpeter (1969; see also Stevenson 1991),  
\be \rstar(\mstar)
  =  2.2\!\times\!10^9 \left({\mstar \over\solarmass}\right)^{-1/3}
\left[{  1 + \left({\mstar \over 0.0032\solarmass}\right)^{-1/2} }\right]^{-4/3} \!\!\!\!\! {\rm cm}
\hspace{0.1truein} (\mstar\!\lesssim\!0.09) ~. \ee

\subsection{Cluster Lifetime Against Evaporation and Collisions}
Since the case of a BH relies on limits to the lifetime of a cluster, 
it is better
to err on the side of caution and use only simple physical considerations. 
Thus, we shall examine the cluster lifetime only against the processes of 
evaporation and physical collisions, and do not 
take into account processes which are not yet fully understood, such as
the post-core collapse evolution of a cluster.  We shall
discuss core-collapse in
\S2.4, and show that including it in the analysis would not have made any
qualitative difference in our conclusions.

An upper limit to the lifetime of any bound stellar system is given by its
evaporation time. Evaporation is the inevitable, continuous process where
stars escape from a stellar-dynamical system
due to weak gravitational scattering.
The evaporation time-scale of a cluster which consists of equal mass objects
is $t_{evap}\!\approx\!300\spa t_{rh}$ (Spitzer \& Thuan 1972, Binney \&
Tremaine 1987, hereafter BT87), where
$t_{rh}\! =\! \left[{0.14 N/\ln(0.4 N)}\right](R_h^3/G M)^{1/2}$
is the median relaxation time (Spitzer \& Hart 1971, BT87), 
and $N\!=\!M/\mstar$ is the number of objects 
of mass $\mstar$ comprising the cluster. 
In terms of the cluster's half-mass $M_h$, and half-mass density $\rho_h$,
we obtain for a Plummer model 
\be   t_{evap} ~ \simeq ~ {4.3\!\times\!10^4 \spa (M_h/\mstar) \over
  \ln[0.8 ~\! (M_h/\mstar)] } \spa \left({\rho_h \over 10^8~\! M_\odot \pc^{-3}
}\right)^{-1/2}  \units {\rm yr.}        \label{evap-time}\ee

The other limit on a cluster lifetime comes from the destruction of the
cluster due to physical collisions. 
The characteristic timescale for each star to physically collide with another,
taking gravitational focusing into account, is (BT87) 
\be t_{coll} = \left[{ 16 \pi^{1/2} n \sigma r_{*}^2
\left(1+{G \mstar\over 2\sigma^2 r_{*}}\right) \,}\right]^{-1} \:\:\:\: ,
\label{tcoll-a} \ee
where $n$ is the number density of objects comprising the cluster, 
and $\sigma$ is their velocity dispersion.
The evolution of a galactic nucleus driven by
collisions has been studied by many authors (\eg Spitzer \& Saslaw 1966;
Spitzer \& Stone 1967; Colgate 1967; Begelman \& Rees 1978), and they 
all agree that the cluster evolution should accelerate rapidly 
once the cluster reaches an age of $t_{coll}$. 
By this time, almost every star will have a collision, and many will
have multiple collisions. 
Stellar debris would settle toward the cluster center, accumulating at the
bottom of the cluster potential well, while stars undergo a process of
runaway coalescence, leading to 
a rapid build-up of a very massive object at the cluster center. 
The most likely end product of this process, which
would leave much dark matter behind, is a massive BH.

Substituting $\sigma_0\!\simeq\! [(2\pi/9) G \rho_0 r_c^2]^{1/2} \!=\!
2^{-1/4} G_{}^{1/2} M_h^{1/3} \rho_h^{1/6}$
for the central velocity dispersion in a Plummer model, 
and $n=\rho_h/\mstar$, equation (\ref{tcoll-a}) yields
\be t_{coll}(\mstar,\rstar)~ = \left[{ 23.8~\! G^{1/2} M_h^{1/3} \rho_h^{7/6} 
\left({r_{*}^2 \over \mstar}\right) 
\left(1 + {\mstar\over 2^{1/2} \rho_h^{1/3} M_h^{2/3} r_{*}}\right) 
\,}\right]^{-1} \:\:\:\: {\rm sec}.  \label{tcoll_2} \ee
The upper limit on the lifetime of a cluster with a given $M_h$ and $\rho_h$, 
which consists of objects of mass $\mstar$ and radius $\rstar$, is then
\be \tau(\rstar,\mstar)~=~{\rm min} \left[{t_{coll},t_{evap}}\right] 
\period  \label{tau-star} \ee
For every combination of $M_h$ and $\rho_h$, we examined the entire 
mass range of every class of non-luminous objects described in \S2.1, 
and found the maximum possible cluster lifetime $\tau_{max}$, where
\be \tau_{max}(M_h,\rho_h) ~=~ {\rm max} \left[{\tau(\rstar,\mstar)}\right] 
\period \label{tmax} \ee
 
Figure 1 presents $\tau_{max}$ for the mass and density ranges found in
galactic nuclei.  For example, the maximum possible
lifetime of a cluster with half mass
of $10^{7}\solarmass$ and half-mass density of $10^{9}\solarmass\pc^{-3}$ is 
$\approx\!10^{10}$ years.
We note that for different combinations of cluster mass and density, the
cluster lifetime peaks at a different stellar type and stellar mass (\eg
$0.6\solarmass$ white dwarfs for $[M_h,\rho_h]\!=\![10^6,10^{12}]$, 
$1.4\solarmass$ neutron stars for $[10^8,10^{12}]$, $1.1\solarmass$ white
dwarfs for $[10^9,10^8]$, 
and $0.004\solarmass$ brown dwarfs for $[10^6,10^5]$). In most parts of the 
examined parameter space, 
the least well-constrained stellar types and masses
are $0.1\hbox{--}1\solarmass$ white dwarfs, and $1.4\solarmass$ neutron
stars.
Note that $\tau_{max}$ increases with mass, which suggests 
that it should be easier to make a case for less massive BHs than for
heavier ones. However, it is generally more difficult to place a stringent
constraint on a cluster density in the case of lower mass objects 
due to the limited angular resolution of the observations.

\subsection{Implications For The Observed Dark Objects}
Observations provide dynamical evidence for large amounts of
dark mass within small regions at the centers of several galaxies.
The size of these regions is usually determined either by 
the angular resolution 
of the observations, or by the inner edge of an observed rotating disk.
The radius of that region provides the most conservative estimate for 
the half-mass radius of a central dark cluster; assuming smaller scales 
for $R_h$ would imply higher densities and thus more rapid cluster 
evolution.  Therefore, we identify the detected mass and mean mass density 
within that central region as $M_h$ and $\rho_h$ of the hypothetical cluster, 
respectively.  Figure 1 presents the current data for the observed BH
candidates (Table 1). 

We find that the lifetime of an hypothetical central cluster must be much 
shorter than the galaxy age only in the cases of NGC 4258 and the Galaxy, 
thus strongly arguing for a point mass.  In all other galaxies, we currently
cannot completely
rule out the possibility of a central dark cluster. It is interesting
to notice that in M87 for example, which contains the most massive dark object 
yet detected ($M_\bullet\!\approx\!3\!\times\!10^9\solarmass$), 
the dynamical constraints 
on a central cluster are very weak. It would require observations with 
nearly three orders of magnitude better angular resolution in order to 
raise the limit on the central density to a point where a dark cluster is 
safely ruled out in that galaxy.  
On the other hand, an improvement of less than one order of 
magnitude in resolution would enable to confidently exclude a dark cluster
in M32, assuming that the inferred amount of dark mass within 
the unresolved central region in that galaxy does not 
drop significantly with increasing resolution.

\subsection{Core-Collapsed Dark Clusters}
Dark clusters must undergo core collapse at a finite age, during which
the core radius shrinks almost to zero, and the central density increases 
enormously. For a Plummer model of equal-mass objects,
core collapse will occur at $t\!\approx\!16t_{rh}$ (Cohen 1980; BT87;
Quinlan 1996), where $t_{rh}$ is the median relaxation time defined in \S2.2.
Core collapse is not necessarily catastrophic for the cluster as a whole, but
it is certainly possible that it may lead to the formation of a BH at the 
cluster center. However, since the mass enclosed within the core drops 
significantly during the collapse, the BH mass will be a very small fraction 
of the dark cluster mass.  
The post-collapse evolution of a cluster, and the growth rate of a seed 
BH depend on complicated processes such as binary interactions and stellar 
mass loss (\eg Ostriker 1985; Goodman 1993). Since the effect of these
processes on the post-collapse cluster evolution are not yet 
fully-understood, we could not include the timescale for core collapse 
as a limit on the cluster lifetime.

Yet, let us suppose that future investigations would reveal that a BH
which contains a significant fraction of the cluster mass must form shortly 
after a core collapses.
Since $16t_{rh}$ is shorter than $t_{evap}$ by a factor of $\sim\!20$
(Eq.~[\ref{evap-time}]), we can expect $\tau_{max}$ (Eq.[\ref{tmax}])
to be shorter by up to the same factor, depending exactly on whether 
the cluster lifetime is more strongly constrained by collisions or 
by evaporation.
Replacing $t_{evap}$ by $16t_{rh}$ in equation (\ref{tau-star}),
we find that $\tau_{max}$ drops by a factor of nearly twenty in the case 
of M87, but it decreases only by a factor of four to $9\!\times\!10^{10}$
and $3\!\times\!10^{10}$ years in the cases of M31 and M32, respectively. 
The latter
result is consistent with the findings of previous investigations of
the central objects in M31 and M32 (Goodman \& Lee 1989; Richstone, Bower 
\& Dressler 1990).  In any case, we see that it is possible that all 
the hypothetical dark clusters which are located below the $10\Gyr$ curve 
in figure 1 have not yet undergone core collapse.  

\section {DISCUSSION}

The main results of this investigation are summarized in the abstract, so we 
avoid redundancy here.
We note that there are two exotic alternatives to a massive BH which cannot
be safely ruled out even in the cases of NGC~4258 and the Galaxy: 
clusters of elementary particles and very low-mass BHs. 
Physical collisions do not affect the evolution of a cluster which consists 
of BHs, and the evaporation timescale can be made arbitrarily
long by giving the BHs an arbitrarily small mass. The lifetime of 
such clusters could exceed $10\Gyr$ if $m_{BH}\!<\!0.04\solarmass$ in the case
of NGC~4258, and if $m_{BH}\!<\!0.005\solarmass$ in the case of the Galaxy
(Eq.~[\ref{evap-time}]).
The most serious difficulty with such an alternative to a massive BH is that 
low mass BHs cannot form in stellar evolution.  They may form
in the early universe, assuming inflationary cosmology (Garcia-Bellido,
Linde \& Wands 1996; Yokoyama 1997).
We also note that the mass range of 
$10^{-8}\!\lesssim\!m_{BH}\!\lesssim\!0.03\solarmass$ BHs is ruled out by 
gravitational microlensing experiments (Alcock \etal 1997), if 
we assume that the same hypothetical BH population comprises both the central 
dark cluster and the Galactic dark halo.

Collisions and evaporation arguments cannot rule out a massive cluster of 
elementary particles either. We can only note that
particles which may comprise a dark cluster cannot be muon or
electron neutrinos of non-zero rest mass, or any other hypothetical 
non-interacting Maxwell-Boltzmann particles, since their fine-grained 
phase-space density would be enormously higher than that allowed by 
cosmological models (see Tremaine \& Gunn 1979). These particles
could be bosons, for example, where the equilibrium phase-space 
density does not have a maximum.
The most serious difficulty with such an alternative to a massive BH is that 
it is very difficult to imagine a process of (inverse) mass segregation,
where $10^{6}\hbox{--}10^{9.5}\solarmass$ of {\it collisionless\/} gas of 
elementary particles could dissipate a large fraction of its energy, and 
evolve toward an extremely dense configuration.
We conclude that clusters of very low-mass BHs and elementary particles 
cannot be ruled out, but their existence is highly implausible. 

Finally, we note that (i) better theoretical understanding of cluster
evolution may enable in the future to tighten some of the limits derived in 
this paper; (ii) the constraint on the density of a dark object at the 
center of our galaxy may be significantly stronger than that used in the
present study (see Genzel \etal 1997); and (iii) all the arguments presented 
in this investigation rely on the assumption that we do not live in a very 
special epoch. 

\acknowledgments
I thank Reinhard Genzel for stimulating discussions, and John Kormendy for
useful comments.

\newpage
\centerline{\bf FIGURE CAPTIONS}
\vspace{0.4truein}

Figure 1 -
The maximum possible lifetime of a dark cluster with half-mass $M_h$ and
half-mass density $\rho_h$, against the processes of evaporation and 
destruction due to physical collisions. 
These clusters may consist of any plausible form of 
non-luminous objects (\S2.1).
The points present the current data for most of the observed BH
candidates (Table 1).
We see that the lifetime of such hypothetical clusters must be much 
shorter than $10\Gyr$ only in the cases of NGC 4258 and our Galaxy, 
thus strongly arguing for a point mass.  
In the case of M87, for example, which contains the most massive central
object yet detected, the dynamical constraints on alternatives to 
a BH are very weak  (see text).

\newpage

\newpage
\centerline{{\bf TABLE 1}}
\vspace{0.4truein}
 
\begin{tabular}{lccc}  \hline\hline
Galaxy & $~~~~~M_h~[\solarmass]~~$  & $~~~~~R_h$~[pc]~~ &
$~~~~~\rho_h~[\solarmass \pc^{-3}]~~$ \\
\hline
NGC~3115   & $1\!\times\!10^9$ &  2   &  $3\!\times\!10^{7}$ \\
NGC~3377   & $1.2\!\times\!10^8$ &  9   &  $4\!\times\!10^{4}$ \\
NGC~4258   & $1.7\!\times\!10^7$ &  0.016  &  $1\!\times\!10^{12}$ \\
NGC~4261   & $2.5\!\times\!10^8$ &  14  &  $2.2\!\times\!10^{4}$ \\
NGC~4342   & $2.5\!\times\!10^8$ & 18   &  $1\!\times\!10^{4}$ \\
NGC~4486B  & $3\!\times\!10^8$ & 17   &  $1.5\!\times\!10^{4}$ \\
NGC~4594   & $5\!\times\!10^8$ &  4.5   &  $1.3\!\times\!10^{6}$ \\
M31        & $1.5\!\times\!10^7$ &  0.6  &  $1.6\!\times\!10^{7}$ \\
M32        & $1.7\!\times\!10^6$ & 0.26   &  $2.3\!\times\!10^{7}$ \\
M84        & $8\!\times\!10^8$ &  8   &  $3.7\!\times\!10^{5}$ \\
M87        & $1.6\!\times\!10^9$ & 3.5   &  $9\!\times\!10^{6}$ \\
The Galaxy & $1.3\!\times\!10^6$ &  0.008  &  $6\!\times\!10^{11}$ \\ \hline
\end{tabular}
 
\vspace{0.5truein}
Table 1 - The half-mass, $M_h\!=\!M_\bullet/2$, of an hypothetical dark
cluster, the current upper limit to its half-mass radius, $R_h$, and the lower
limit to its half-mass density, $\rho_h\!=\!(3M_h/4\pi R_h^3)$,
for the observed BH candidates:
NGC~3115 (Kormendy \etal 1996a),
NGC~3377 (Kormendy \etal 1998),
NGC~4258 (Maoz 1995),
NGC~4261 (Ferrarese, Ford \& Jaffe 1996),
NGC~4342 (van den Bosch \& Jaffe 1997),
NGC~4486B (Kormendy \etal 1997),
NGC~4594 (Kormendy \etal 1996b),
M31 (KR95),
M32 (van der Marel \etal 1997),
M84 (Bower \etal 1997),
and M87 (Marconi \etal 1997).
The limit on $\rho_h$ at the center of our galaxy is based on the nearest star
to SgA$^{*}$ with a measured proper motion (Eckart \& Genzel 1997).

\newpage
\begin{figure}[ht]
\textwidth=3in
 \centerline{\psfig{figure=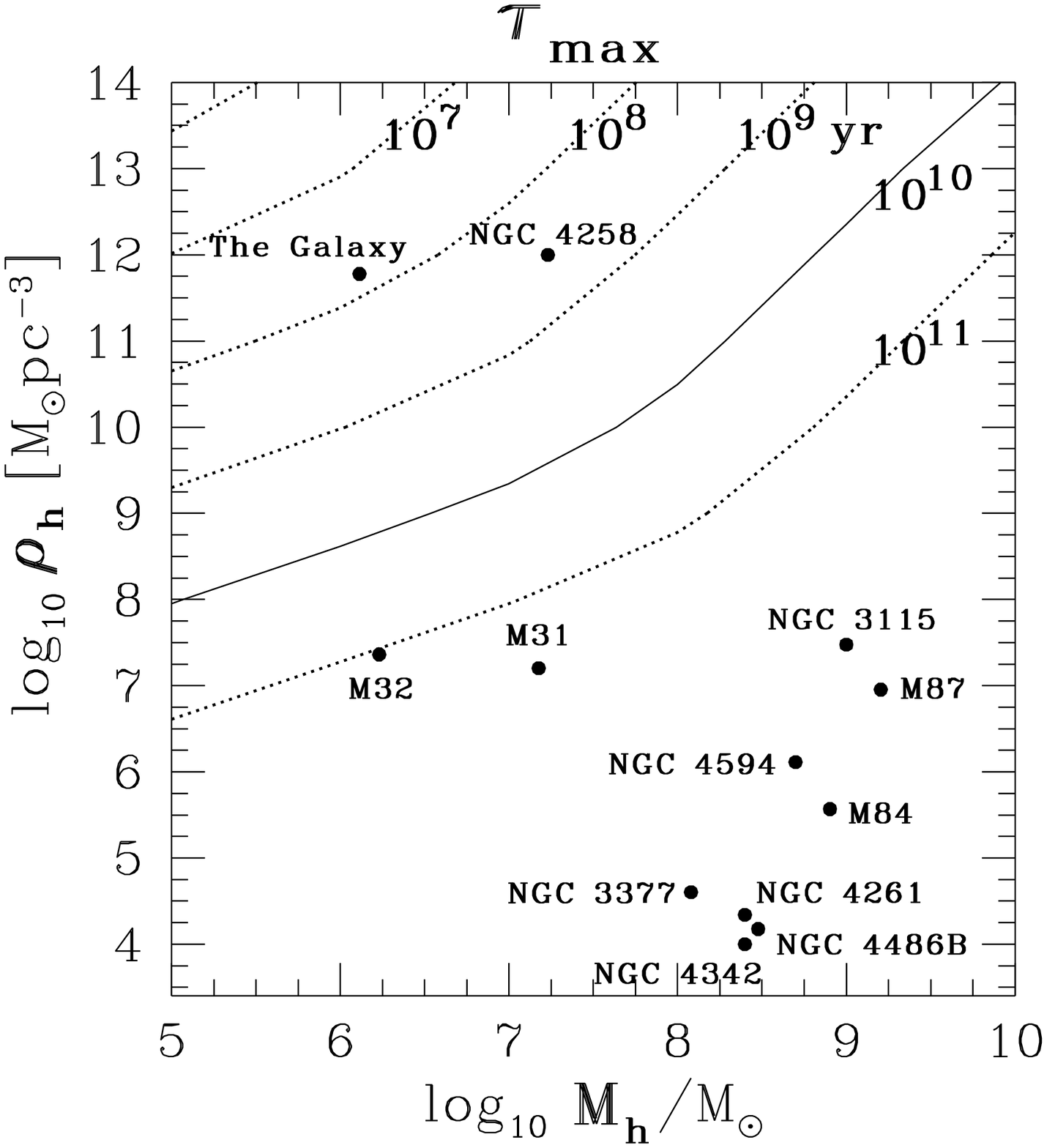,height=7.0in,width=7.0in}}
\end{figure}


\begin{references}
\reference{} Alcock, C., \etal 1997, \apj, 486, 697 
\reference{} Bower, G., \etal 1997, \apjl, in press
\reference{} Garcia-Bellido, J., Linde, A. \& Wands, D. 1996, \prd,
   54, 6040
\reference{} Binney, J., \& Tremaine, S. 1987, {\it Galactic Dynamics}, 
   Princeton: Princeton University Press ~(BT87)
\reference{} Cohen, H. 1980, \apj, 242, 765
\reference{} Colgate, S.A. 1967, \apj, 150, 163
\reference{} Eckart, A., \& Genzel, R. 1997, \mnras, 284, 576
\reference{} Ferrarese, L., Ford, H.C. \& Jaffe, W. 1996, \apj, 470, 444
\reference{} Genzel, R., Eckart, A. Ott, T. \& Eisenhauer, F. 1997, \mnras,
   in press.
\reference{} Goodman, J. 1993, In {\it Structure and Dynamics of Globular
   Clusters}, eds. S.G.~Djorgovski \& G.~Meylan, ASP Conferences Series,
   Vol.~50, p.~87
\reference{} Goodman, J. \& Mok Lee, H. 1989, \apj, 337, 84
\reference{} Kormendy, J., \& Richstone, D. 1995, \araa, 33, 581 ~(KR95)
\reference{} Kormendy, J. \etal 1996a, \apj, 459, L57
\reference{} Kormendy, J. \etal 1996b, \apj, 473, L91
\reference{} Kormendy, J. \etal 1997,  \apj, 482, L139
\reference{} Kormendy, J., Bender, R., Evans, A.S. \& Richstone, D. 1998, to 
   appear in the \aj.  
\reference{} Maoz, E. 1995, \apj, 447, L91
\reference{} Marconi, A., Axon, D.J., Macchetto, F.D., Capetti, A.,
  Sparks, W.B. \& Crane, P. 1997, to appear in \mnras
\reference{} Nauenberg, M. 1972, \apj, 175, 417
\reference{} Ostriker, J.P. 1985, in {\it Dynamics of Star Clusters}, IAU
   Symp.~No.~113, eds.~J.~Goodman \& P.~Hut, Dordrecht: Reidel, p.~347 
\reference{} Quinlan, G.D. 1996, {\it New Astron.}, 1, 255
\reference{} Richstone, D., Bower, G., \& Dressler, A. 1990, \apj, 353, 118
\reference{} Spitzer, L.,  \& Saslaw, W.C. 1966, \apj, 143, 400
\reference{} Spitzer, L.,  \& Stone, M.E. 1967, \apj, 147, 519
\reference{} Spitzer, L.,  \& Hart, M.H. 1971, \apj, 164, 399
\reference{} Spitzer, L.,  \& Thuan, T.X. 1972, \apj, 175, 31
\reference{} Stevenson, D.J. 1991, \araa, 29, 163
\reference{} Tremaine, S. \& Gunn, J. 1979, \prl, 42, 407
\reference{} van den Bosch, F.C., \& Jaffe, W. 1997, in The Nature of
  Elliptical Galaxies, ed.~M.~Arnaboldi, G.~S.~Da Costa \& P.~Saha,
  (San Francisco: ASP), p.~142
\reference{} van der Marel, R.P., Cretoon, N., de Zeeuw, T., \& Rix, W.-H. 
  1997, submitted to \apj
\reference{} Yokoyama, J. 1997, \aap, 318, 673
\reference{} Zapolsky, H.S., \& Salpeter, E.E. 1969, \apj, 158, 809
\end{references}
\end{document}